\documentclass[twocolumn, superscriptaddress]{revtex4}
\usepackage{graphicx}
\usepackage{bm}
\usepackage{hyperref}
\usepackage{amsmath,amssymb}
\usepackage{amsfonts}
\usepackage{siunitx}
\usepackage[utf8]{inputenc}
\usepackage{float}
\usepackage{soul}
\newcommand{\be}{\begin{equation}}
\newcommand{\ee}{\end{equation}}
\newcommand{\dv}{\,{\rm d}V}

\begin{document}
\title{Defect-sensitive High-frequency Modes in a Three-Dimensional Artificial Magnetic Crystal}
\author{Rajgowrav Cheenikundil}
\affiliation{
 Universit\'e de Strasbourg, CNRS,\\
 Institut de Physique et Chimie des Mat\'eriaux de Strasbourg, UMR 7504, F-67000 Strasbourg, France
}
\author{Massimiliano d'Aquino}
\affiliation{Department of Electrical Engineering and ICT, University of Naples Federico II, Naples, Italy}
\author{Riccardo Hertel}%
 \email{riccardo.hertel@ipcms.unistra.fr}
\affiliation{
 Universit\'e de Strasbourg, CNRS,\\
 Institut de Physique et Chimie des Mat\'eriaux de Strasbourg, UMR 7504, F-67000 Strasbourg, France
}

\begin{abstract}
Modern three-dimensional nanofabrication methods make it possible to generate arbitrarily shaped nanomagnets, including periodic networks of interconnected magnetic nanowires. Structurally similar to optical or acoustic metamaterials, these arrays could represent magnetic variants of such artificial materials.
Using micromagnetic simulations, we investigate a three-dimensional array of interconnected magnetic nanowires with intersection points corresponding to atomic positions of a diamond lattice. The high-frequency excitation spectrum of this artificial magnetic crystal (AMC) is governed by its microstructure and, to a lesser extent, by the magnetic configuration. The magnetic system displays characteristics of three-dimensional artificial spin ice. It can contain Dirac-type magnetic defect structures, which modify the magnonic spectrum of the AMC similarly as defect sites in a natural diamond crystal influence optical absorption spectra. Our study opens new perspectives for applying such materials in high-density magnonic devices and shows that AMCs represent a promising category of magnonic materials with tunable properties.
\end{abstract}

\maketitle

The desire to design metamaterials---synthetically prepared material systems with properties that do not occur naturally---has long inspired condensed-matter physicists \cite{kadic_3d_2019,liu_metamaterials_2011,smith_metamaterials_2004}.  
The unusual effective material parameters of metamaterials arise from their microstructure, consisting of microscopic building blocks repeated periodically in space \cite{valentine_three-dimensional_2008, burckel_micrometer-scale_2010}. 
This periodic arrangement results in an artificial crystal structure, and it affects the material’s photonic, phononic, or electronic response, causing unusual optical~\cite{soukoulis_past_2011}, acoustic~\cite{cummer_controlling_2016}, or electromagnetic~\cite{fan_dynamic_2015} properties. 
Research on this topic has recently experienced a significant boost owing to progress in three-dimensional (3D) fabrication techniques \cite{kadic_3d_2019}. 

Similarly, the advent of three-dimensional nanomagnetic fabrication techniques has sparked considerable interest in the magnetism community \cite{fernandez-pacheco_three-dimensional_2017,fernandez-pacheco_writing_2020,fischer_launching_2020,williams_two-photon_2018,keller_direct-write_2018,teresa_review_2016}. 
Still, these techniques are relatively new, and they have not yet had a strong impact on the design of magnetic metamaterials. 
By harnessing the possibilities provided by FEBID (focussed electron beam induced deposition) it was recently shown that virtually any three-dimensional geometry can be generated \cite{sanz-hernandez_artificial_2020,skoric_layer-by-layer_2020}. The technique allows for a layer-by-layer deposition of magnetic materials with geometries of complex shape and nanoscale feature size. Using this method, it is possible to prepare periodic three-dimensional magnetic arrays of interconnected nanowires \cite{da_camara_santa_clara_gomes_magnetic_2016,burks_3d_2021} so that their vertex positions correspond to atomic sites of molecules or crystals, with the nanowires connecting the nearest-neighbor sites \cite{keller_direct-write_2018}. The resulting artificial magnetic crystals (AMCs) are a largely unexplored category of magnetic nanostructures, especially their high-frequency magnetic properties \cite{sahoo_observation_2021}.

Such periodic 3D arrays of interconnected nanowires could constitute magnetic metamaterials, with physical properties imprinted by their geometry and different from those of the material in its natural form. In analogy to optical metamaterials \cite{valentine_three-dimensional_2008}, whose electromagnetic spectrum is governed by the microstructure, AMCs may be designed so that they represent a medium in which magnons, the fundamental excitations in magnetism, display similarly unusual and microstructure-driven properties.  
If successful, the development of AMCs with tunable magnonic properties could conduce to a departure from the commonly used approach of designing {\em magnonic} crystals~\cite{krawczyk_review_2014,chumak_magnonic_2017}.  In their traditional form, magnonic crystals are arrays of spatially separated micromagnets arranged on a periodic grid, which---through their dipolar coupling---display collective features of the high-frequency magnetic excitations. Three-dimensional AMCs formed by interconnected unit cells could represent an advanced type of magnonic crystal in the form of bulk media with nanoscale microstructure and tunable magnonic properties.

The interconnected nanowire arrays forming an AMC constitute three-dimensional versions of artificial spin ice (ASI) systems in terms of their micromagnetic properties~\cite{mistonov_three-dimensional_2013,koraltan_tension-free_2021,may_magnetic_2021}.
In the past decade, ASI lattices have advanced to an essential research topic in magnetism~\cite{nisoli_colloquium_2013,skjaervo_advances_2020,heyderman_artificial_2013,wang_artificial_2006,zhang_crystallites_2013}. In their traditional form, these systems consist of sub-micron-sized patterned magnetic elements with single-domain magnetization arranged in a regular grid. 
The magnetic state of the ASI is characterized by the magnetic configurations at the crossing points of the array (the vertex sites) \cite{morrison_unhappy_2013}. Vertices in which the surrounding magnetic configuration results in a non-zero net magnetic flux toward or away from them are magnetically frustrated defect sites with a monopole-type emergent charge~\cite{mengotti_real-space_2011,hugli_artificial_2012,gilbert_emergent_2014,ladak_direct_2010}. Numerous magnetic defects, which appear in different forms~\cite{gilbert_emergent_2014}, can be distributed onto the vertex sites throughout the ASI, resulting in a quasi-continuum of possible magnetic configurations \cite{nisoli_ground_2007,perrin_extensive_2016}. From a fundamental perspective, research on ASI systems is driven by the interest to explore their rich and complex features. In addition, from an application point of view, ASIs are promising candidates for future components in low-energy programmable magnonic devices \cite{lenk_building_2011,barman_2021_2021}, given that their high-frequency oscillations can be manipulated by changing the array's magnetic configuration \cite{krawczyk_review_2014,gliga_spectral_2013,cheenikundil_switchable_2021,frotanpour_vertex_2021,
arroo_sculpting_2019,lendinez_magnetization_2019,bhat_magnetization_2016}.

While ASIs have mainly been studied in the form of separated magnetic elements \cite{zhang_crystallites_2013,wang_artificial_2006,farhan_direct_2013}, recent studies have extended the concept to interconnected nanowire arrays \cite{frotanpour_vertex_2021,shi_kerr_2020,saavedra_dynamic_2020,sun_magnetization_2017}. 
Moreover, the aforementioned three-dimensional nanomagnetic fabrication techniques have made it possible to study {\em three-dimensional} arrays of interconnected magnetic nanowires \cite{may_magnetic_2021,sahoo_observation_2021,fischer_launching_2020}. Using micromagnetic finite-element simulations, we investigate the magnetic structure as well as the high-frequency magnetization dynamics in a mesoscopic 3D array of nanowires forming a diamond-type artificial crystal. Our study shows that, similar to the effect of color centers in an actual diamond \cite{maki_properties_2009,aharonovich_diamond_2014,bradac_quantum_2019}, magnetic defect structures in 3D AMCs generate pronounced peaks in the spectrum of magnetic modes, thereby leading to an effective ``coloring'' of the magnetic high-frequency spectrum. Besides the appealing physical analogy between the impact of these magnetic defects on the AMCs' magnonic properties and those of real defects on the optical properties of natural crystals, these findings could have implications for the development of magnonic metamaterials.

The dependence of the AMC’s spectrum on the density and type of magnetic defect states discussed here opens up additional possibilities---unavailable in more traditional metamaterial types---of manipulating the system's high-frequency properties. A control of this type is in agreement with the general goal of research in metamaterials, which is to tailor the effective material parameters of synthetic crystals by controlling the distribution of only a few elementary ingredients within the building blocks. In their recent review, Kadic et al.~\cite{kadic_3d_2019} have compared this approach to the ability of a graphical printer to generate a vast multitude of colors from just three color cartridges.  
In the context of magnetic metamaterials, the insertion of a defect site in the AMC can be interpreted as the replacement of one type of unit cell with another.
Leveraging on the detailed knowledge acquired in the magnetism community on the properties of ASI and magnetic defect structures, our study provides new perspectives on the development of magnetic metamaterials with potential applications in devices for storage, magnonic operations, and neuromorphic computations.

\section{Model system}
\begin{figure}[h]
\includegraphics[width=\linewidth]{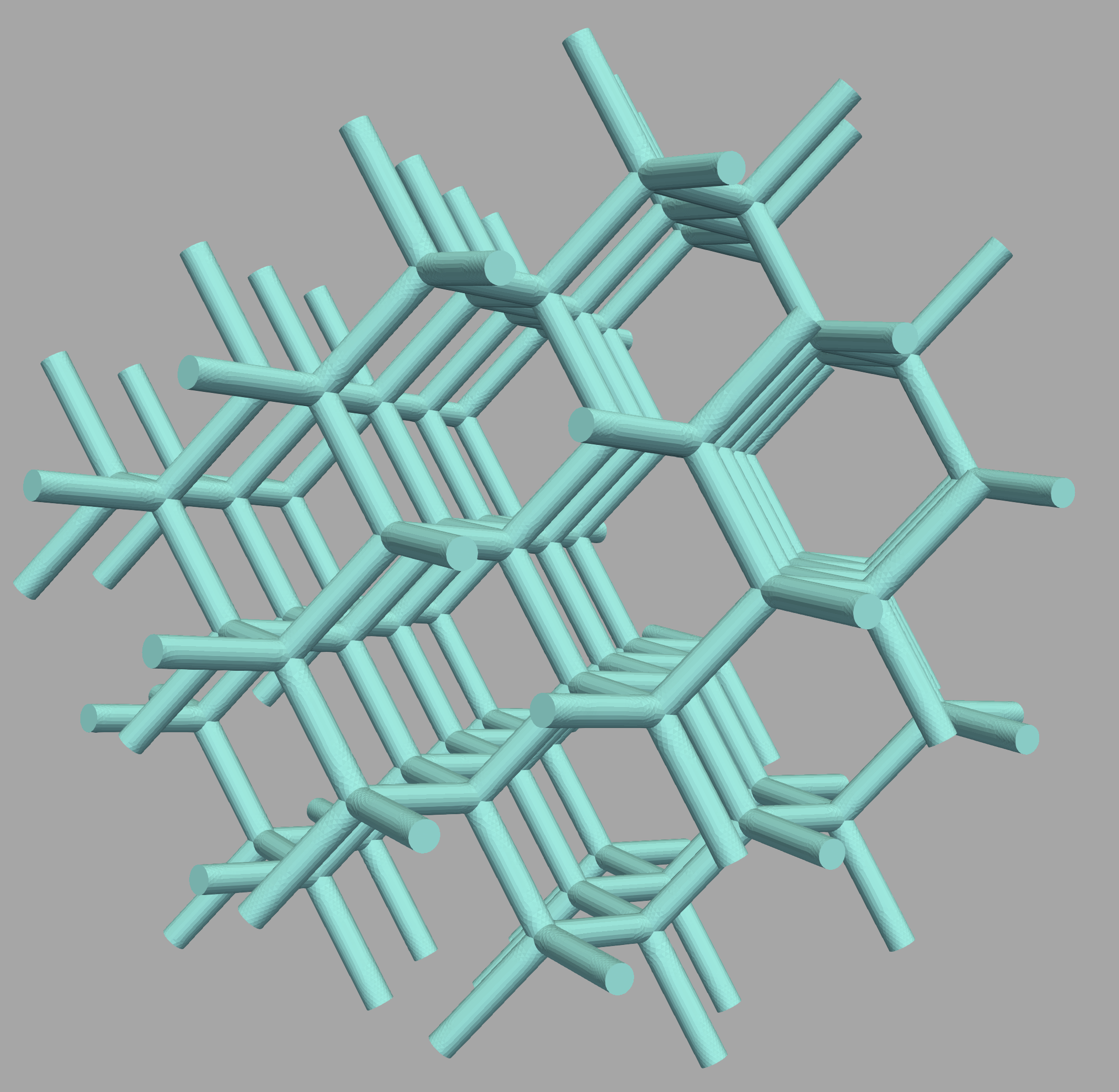}
\caption{Three-dimensional nanowire array with a diamond-type structure. The array contains 238 nanowires and 155 vertices. Each wire has a length of \SI{70}{\nano\meter} and a diameter of \SI{12}{\nano\meter}. We assume ideally soft-magnetic material with properties corresponding to those of typical FEBID-deposited nanoarchitectures \cite{fischer_launching_2020}.
	\label{fig:diamond}}
\end{figure}

Our model of the AMC structure, illustrated in Fig.~\ref{fig:diamond}, consists of cylindrical nanowires  connecting nearest-neighbor sites of a diamond lattice. Each nanowire has a length of $l=\SI{70}{\nano\meter}$ and a diameter $d=\SI{14}{\nano\meter}$), resulting in a diamond-type AMC structure with a cubic lattice constant of about \SI{162}{\nano\meter}. 

 The arrangement of the magnetization forming in the artificial crystal has the characteristics of a three-dimensional ASI, since each wire is magnetized homogeneously along its axis. This combination of two branches of research in magnetism---3D nanomagnetism and ASI---was recently discussed on the example of a buckyball-type geometry \cite{cheenikundil_switchable_2021}. It is well known that the magnetization state of ASIs is not unique \cite{nisoli_ground_2007,montaigne_size_2014}. The lowest-energy arrangement of the magnetization is a defect-free magnetic configuration that satisfies the so-called ice rule at every vertex \cite{gilbert_emergent_2014}, but ASI lattices can also contain several magnetic defect structures carrying monopole-type magnetostatic charges \cite{montaigne_size_2014,nisoli_colloquium_2013,morrison_unhappy_2013}.

\section{Vertex charges and static configuration}
Before addressing the magnetic high-frequency modes of the three-dimensional array, it is instructive to analyze the possible micromagnetic configurations at the vertex sites, since---as we shall see---they have a decisive impact on the array's dynamic properties.
The diamond-type AMC has a coordination number of four: each vertex inside the 3D lattice is the intersection point of four nanowires. Regarding the coordination number, this type of lattice is therefore comparable to the well-known two-dimensional square ASI lattice \cite{wang_artificial_2006, porro_exploring_2013,gliga_spectral_2013,remhof_magnetostatic_2008}, which can be considered to be its two-dimensional (2D) counterpart. 
The main constituents of an ASI---in this case the cylindrical nanowires---are usually of Ising type, meaning that their magnetic structure is homogeneous and aligned along the particle axis. The assembly of such Ising-type nanomagnets into a regular array can generate structures of varying complexity and magnetic frustration at the vertices. Vertex configurations in ASIs can be characterized by their magnetic charge \cite{montaigne_size_2014,cheenikundil_switchable_2021}. 

\begin{figure}[ht]
\includegraphics[width=\linewidth]{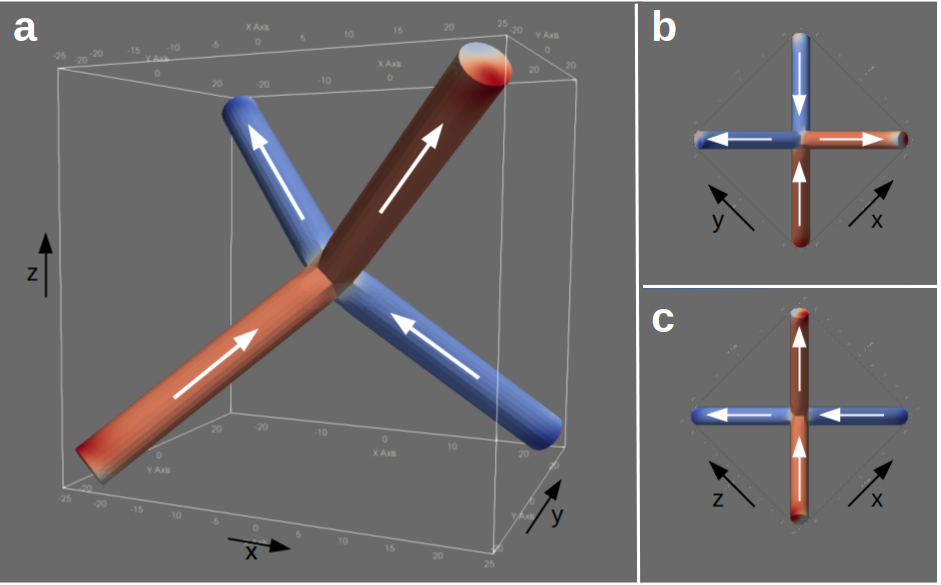}
\caption{Zero-charge vertex configuration in a tetrapod structure---the elementary building block of a diamond-type 3D ASI lattice (a). Only one type of zero-charge configuration exists in 3D, whereas different zero-charge vertex structures are known from 2D square ice systems. The flavors of zero-charge vertex configurations shown in panels (b) and (c) known from 2D square ice ASI lattices differ qualitatively in their micromagnetic structure, and they have different physical properties. Nevertheless, they can be interpreted as projections of a single 3D structure onto different planes. The color code represents the $x$-component of the magnetization.}
	\label{fig:projections}
\end{figure}

The definition of the vertex charge is based on the fact that---assuming Ising-type magnetization in each nanowire---each nanowire carries magnetic flux either toward the vertex or away from it. Any imbalance in the number of wires with in- and out-flowing magnetic flux results in a net magnetic charge at the vertex, akin to a magnetic monopole \cite{mengotti_real-space_2011,gilbert_emergent_2014,ladak_direct_2010}. With this premise, the charge of a vertex with coordination number four can take one of five possible values: $+4q$ (``four-in''), $+2q$ (``three-in/one-out''), 0, (``two-in/two-out''), $-2q$ (``one-in/three-out''), and $-4q$ (``four-out''). The value of the magnetic charge $q$ can be estimated to $q=r^2\pi M_s$, where $r$ is the nanowire radius and $M_s$ is the saturation magnetization; a material-specific constant. Different vertex configurations and their resulting charges have been discussed in detail for 2D square ASI lattices \cite{gilbert_frustration_2016,perrin_extensive_2016}, and the same type of charge-based classification can also be made for vertex configurations in 3D diamond-type lattices. There is, however, a difference regarding the multiplicity of uncharged magnetic structures at the vertices in the 2D as compared to the 3D case. While in 2D lattices the zero-charge configuration exists in two different flavors \cite{gilbert_frustration_2016}, it only appears in one form in the case of 3D diamond-type lattices \footnote{
All zero-charge configurations in the tetrapod structure are equivalent as they can be mapped onto the one shown in Fig.~\ref{fig:projections}a) by means of a rotation and/or a time-inversion operation $\vec{M}\to-\vec{M}$. Such a mapping is not possible for the different 2D variants in a square lattice shown in Fig.~\ref{fig:projections}b) and c).
}, see Fig.~\ref{fig:projections}. This reduction of complexity in the 3D system --compared to its 2D analogue-- simplifies the interpretation of the magnetic configuration in the diamond-type structure as it allows to identify the vertex configurations uniquely by means of their charge, without the need to distinguish different flavors of zero-charge vertices. 

\begin{figure}[ht]
\includegraphics[width=\linewidth]{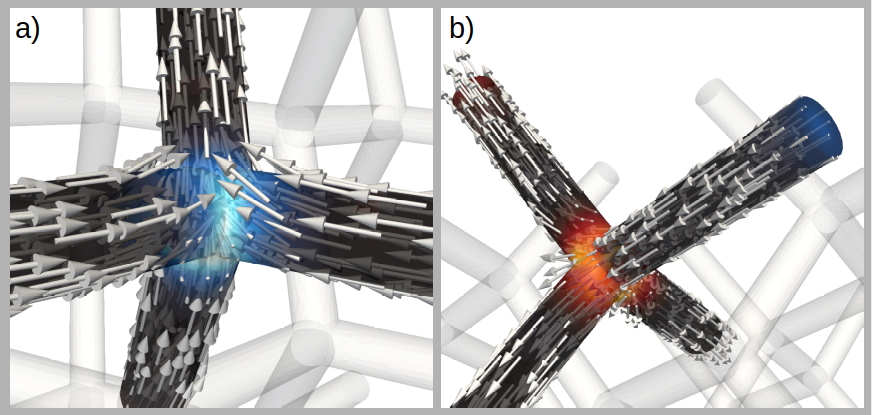}
\caption{Different types of effective magnetic charges in the diamond-type 3D lattice. (a) Double-charge vertex with a ``three-in/one-out'' configuration, leading to an effective charge of $+2$. The orange colored region in panel (b) shows an equivalent ``one-in/three-out'' vertex with charge $-2$. The double-charge vertex in (b) connects two nanowires at the crystal surface whose dangling ends carry an effective charge of $\pm 1$. In both images, the color code displays the divergence of the magnetization. With this representation, the single charges appear as a weak orange and blue contrast near the ends of the wires at the top left and the top right, respectively. For better visibility, only a small subset of the computed magnetic moments is displayed with arrows. \label{fig:chargetypes}}
\end{figure}

Inside the 3D diamond-type ASI, each nanowire is connected to two vertices with even charges ($\pm 4q$, $\pm 2q$, or $0$). In addition to the vertices within the volume, a different type of charge exists at the lattice boundary (``surface''), where the wires are connected only to one vertex. The dangling ends of these outer nanowires can be considered as vertices with coordination number one and charge $\pm 1q$, where the sign indicates whether the magnetization points inwards or outwards. In total, the 3D diamond-type ASI lattice with Ising-type nanowires thus contains vertices of charges $\pm 4q$, $\pm 2q$, $\pm 1q$, and $0$. Owing to time-inversion symmetry, the magnetic configurations of positive and negative charged vertices are equivalent. It is therefore sufficient to distinguish the vertices only by their absolute charge, which effectively reduces the number of vertex configurations to four. 
In the diamond-type ASI lattice that we have studied, the diversity of vertex structures is further diminished by the fact that vertices with quadruple charges were never observed in the simulations, which indicates that these configurations are unstable. The magnetic state of the diamond-type ASI lattice is thus effectively characterized by the spatial distribution of double-charged defects of the type ``three-in/one-out'' and ``one-in/three-out''. A neutrality condition applies to the sum of vertex charges in the lattice, but not to the individual types: the presence of a double-charged vertex does not imply the formation of an additional double-charged vertex of opposite sign, so long as the sum of all vertex charges (including those at the surface) is zero. 

The ice-rule-obeying zero-charge vertices are energetically optimal magnetic arrangements within the lattice, whereas charged vertex configurations are localized magnetic defect structures increasing both the exchange and the magnetostatic energy. 
Defect charges within the AMC increase the magnetic disorder, which can be quantified by the density of charged vertices. 
In spite of their heightened energy compared to the defect-free state, magnetic structures containing several charged vertices can develop as metastable states, and they may persist even at sizeable external fields. Such multiplicity of metastable magnetization states is well-known \cite{nisoli_colloquium_2013,perrin_extensive_2016} from 2D ASI systems---a behavior that is found also in our 3D diamond-type lattice. Removing defect configurations from the system requires external fields strong enough to magnetically saturate the sample, which is only achieved at fields exceeding about \SI{100}{\milli\tesla}. 

\section{Magnonic spectrum}
To study the  magnonic spectrum of the diamond-type AMC and how it is impacted by magnetic defects, we consider two different remanent magnetic configurations: a defect-free state, which will serve as reference state, and a meta-stable state with random disorder. 
In the defect-free state, all vertices within the volume have the same two-in/two-out--type zero-charge configuration. Since the magnetic structure of the defect-free AMC is identical for each unit cell, the magnetic state shares the same periodicity of the crystal structure. The disordered state, on the other hand, contains several vertices with double charge configurations (three-in/one-out or one-in/three-out), and its magnetic structure does not display any spatial periodicity. Double-charge vertices of different sign are randomly scattered within the AMC. In our case, the disordered state contains 35 double-charge vertices and 48 zero-charge vertices. This disordered state is only one realization out of a quasi-continuum of possible configurations 
\cite{nisoli_ground_2007,morgan_thermal_2011,moller_magnetic_2009}.

\begin{figure}[ht]
\includegraphics[width=\linewidth]{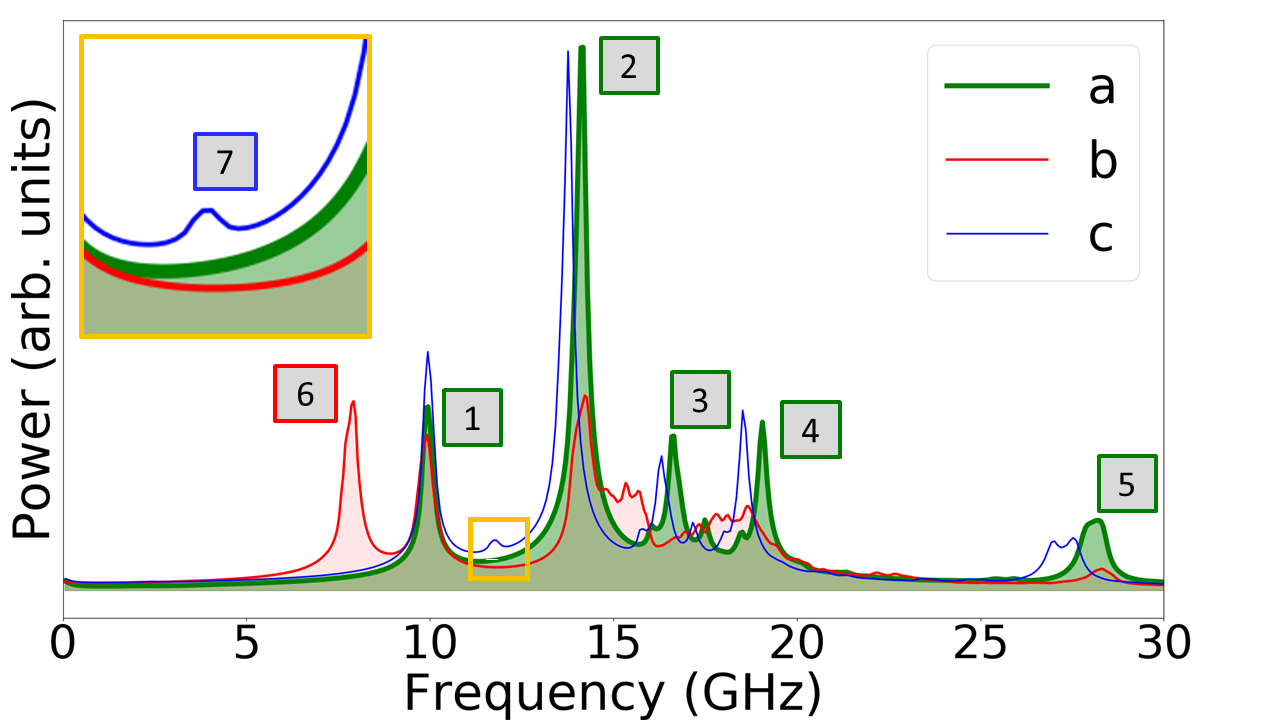}
\caption{Magnetic excitation spectrum of the ordered magnetic crystal and defect-induced modifications.
The AMC displays characteristic frequencies at which the magnetization oscillates. The spectrum (a) of the ordered state contains five distinct peaks, each corresponding to a different magnetic excitation. The magnetically disordered state yields a spectrum (b) with an additional low-frequency mode \fbox{6}. The spectrum (c) refers to an AMC containing a structural defect site. 
	\label{fig:defectFreeSpectrum}}
\end{figure}

\begin{figure*}[ht]
\includegraphics[width=\linewidth]{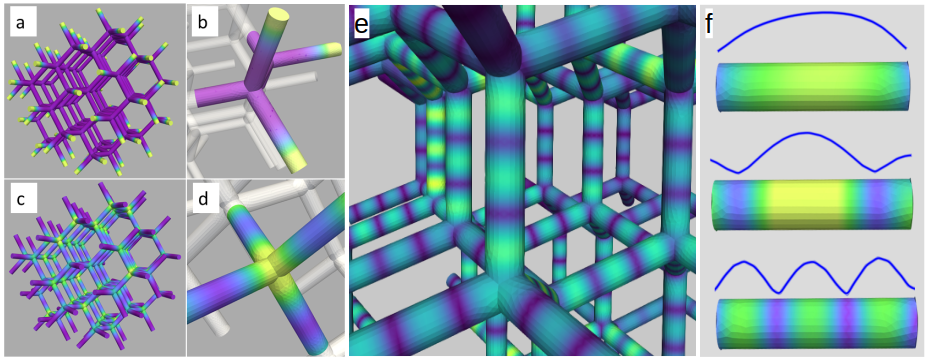}
\caption{Fundamental modes of the defect-free diamond-type AMC. The low-frequency modes, corresponding to the peaks labeled \fbox{1} and \fbox{2} in Fig.~\ref{fig:defectFreeSpectrum}, are oscillations localized at the surface and volume vertices, respectively.(a) The mode at \SI{9.9}{\giga\hertz} refers to oscillations of the free ends, i.e, to the outermost vertices at the crystal surface, as shown in the magnified view in frame (b). The strongest excitation in the spectrum, displayed in frame (c), is due to vertices within the volume which become active at \SI{14.2}{\giga\hertz}. The magnified view (d) shows the localization of the mode at the vertex sites. (e) The high-frequency modes at \SI{16.7}{\giga\hertz}, \SI{19.1}{\giga\hertz}, and \SI{28.2}{\giga\hertz} (corresponding to the peaks \fbox{3}, \fbox{4}, and \fbox{5}, respectively) are oscillations describing longitudinal standing spin waves in the nanowires. Frame (f) shows the computed mode profile along the nanowires for these standing-wave modes, with frequency increasing from the top to the bottom.  In all panels, the color code ranging from purple to yellow describes regions of minimum and maximum oscillation amplitude, respectively.
	\label{fig:vertexModesDefectFree}}
\end{figure*}

Let us first analyze the high-frequency modes developing in the ordered configuration. 
The magnonic spectrum of the defect-free configuration, shown in green (line a) in Fig.~\ref{fig:defectFreeSpectrum}, displays five clearly defined peaks corresponding to distinct magnetic modes within the ASI lattice. 
Fourier analysis methods, through which the spatial distribution of the local oscillation amplitude (i.e., the mode profile) can be extracted for each frequency, reveal the central role played by the geometry in each of the peaks in the spectrum. Each of the prominent modes can be attributed to oscillations localized at different geometric constituents of the AMC: the vertices in the volume, the cylindrical nanowires, and the wire ends at the crystal surface.

The lowest-frequency mode at \SI{9.9}{\giga\hertz}, denoted with the symbol \fbox{1} in Fig.~\ref{fig:defectFreeSpectrum}, is a surface contribution arising from the oscillations of the single-charge vertices on the outer shell. The color-code through which the mode profile is represented in Fig.~\ref{fig:vertexModesDefectFree}(a) and (b) shows inactive regions in purple color, while the yellow color indicates regions of maximum oscillation amplitude. Only the free ends of the single-connected wires partake in the magnetic oscillation at this frequency. This is the only surface contribution in the magnonic spectrum of the AMC. The relative height of the peak \fbox{1} with respect to the other modes therefore depends on the crystal size and the resulting surface-to-volume ratio. The oscillations at the wire ends appear to be decoupled from one another. No clear correlation is observed among the phases of this oscillation at different sites (see Movie 1 in the supplemental material).

The most pronounced peak in the spectrum develops at \SI{14.1}{\giga\hertz} (see label \fbox{2} in Fig.~\ref{fig:defectFreeSpectrum}). The corresponding mode is an oscillation of the magnetization at the zero-charge vertices within the array. The profile of this mode, shown in Fig.~\ref{fig:vertexModesDefectFree}c) and d), displays a clear localization at the center of the tetrapod-shaped vertices. Only the vertices within the volume are active; the free ends do not oscillate at this frequency. 
Contrary to the mode localized at the free ends at the surface, the oscillation phases of the individual vertices is correlated and almost synchronous throughout the volume of the lattice (see Movie 2 in the supplemental material). 

Three distinct high-frequency peaks can be identified in the spectrum of Fig.~\ref{fig:defectFreeSpectrum}, labeled \fbox{3}, \fbox{4}, and \fbox{5}. At these frequencies, standing waves develop in the nanowires. Fig.~\ref{fig:vertexModesDefectFree}e) shows the mode at \SI{28.2}{\giga\hertz}, in which the magnetic oscillation within each wire contains four nodes and three anti-nodes. The profiles of this category of modes are displayed in Fig.~\ref{fig:vertexModesDefectFree}f). In that image, the wires have been isolated graphically from the main crystal for visualization purposes, but they are magnetically and structurally embedded into the AMC. In these standing wave modes, remarkably, the wire ends (i.e., the vertices) neither act as free nor as fixed ends of the oscillation. At \SI{16.7}{\giga\hertz} and \SI{28.2}{\giga\hertz}, the vertices remain inactive, while at \SI{19.1}{\giga\hertz} the magnetization also oscillates at the vertices. Although there appears to be a certain phase correlation of the standing-wave modes in different wires within the crystal, it is not always as clear as in the case of the vertex mode \fbox{2}. The modes can be viewed in the movies 3, 4, and 5 of the supplemental material. A clear macroscopic correlation can be observed only for mode \fbox{4}, which is the only one at which the volume vertices participate in the oscillation. This suggests that long-range dynamic correlations primarily result from the dynamic interaction between the volume vertices.

\section{Modes in the magnetically disordered state}
Up to this point, we have investigated the magnetic high-frequency modes of an AMC in which the structure is magnetically ordered, such that all vertices in the volume have a zero-charge magnetic configuration. Although the magnetic structure is inhomogeneous (there are six different magnetization directions in the nanowires, according to their orientation in space), the magnetic modes do not appear to be influenced by the magnetic structure. The modes discussed so far appear to be tightly connected to the structural or geometric elements of the AMC: vertices, nanowires, and surface; not to the magnetization. 

A strong influence of the magnetic structure on the magnetic modes, however, is revealed by analyzing the magnetically disordered AMC. The excitation spectrum of the disordered AMC, printed in red in Fig.~\ref{fig:defectFreeSpectrum} (line b), displays significant differences with respect to the ordered AMC (line a). The most striking change is the appearance of an additional low-frequency mode \fbox{6} at \SI{7.9}{\giga\hertz}. This mode describes the oscillation of the double-charge vertices in the volume (three-in/one-out and one-in/three-out configurations), which are absent in the ordered state. The appearance of this mode is an example of a high-frequency feature of the AMC that is directly related to its magnetic configuration. The zero-charge vertices \fbox{2}, which are still present in the disordered AMC, oscillate at a frequency markedly different from the oscillation of the double-charge vertices, even though they are localized at identical geometrical sites.

\begin{figure}[ht]
\includegraphics[width=\linewidth]{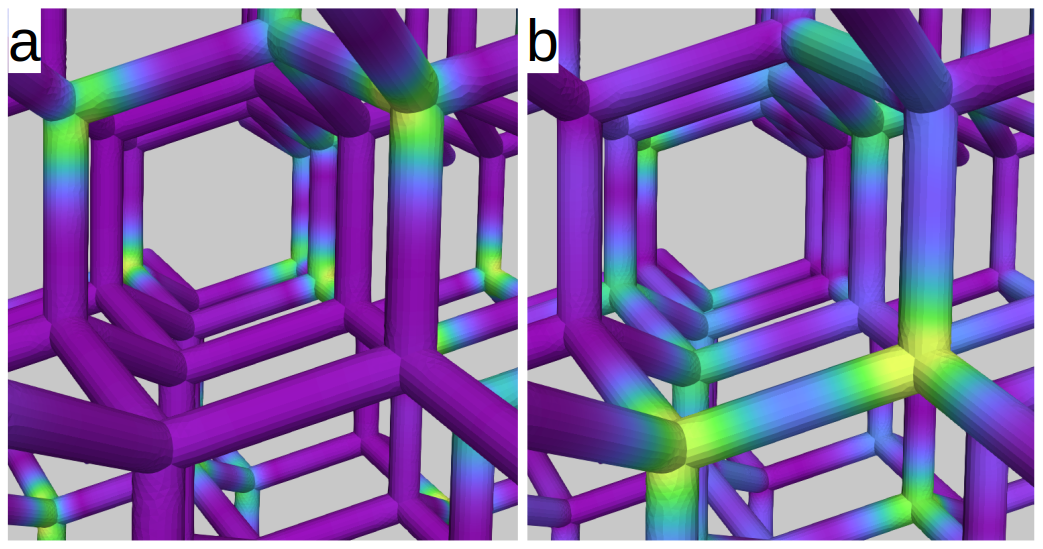}
\caption{Vertex modes in the disordered state. Frame (a) shows the mode profile at \SI{14.2}{\giga\hertz}. Only the vertices with zero-charge type magnetic configuration partake in the oscillation at that frequency. The frequency as well as the local mode profile of the oscillating zero-charge vertices are the same as in the ordered state, shown in Figs.~\ref{fig:vertexModesDefectFree}c,d). The vertices with double charge configuration are active at a significantly lower frequency (\SI{7.9}{\giga\hertz}). The mode profile of this oscillation, shown in frame (b), highlights the activity of a set of vertices which is complementary to that in the mode shown in frame (a).
	\label{fig:zeroAndDoubleChargeModes}}
\end{figure}

The different oscillation profiles of the zero-charge and double-charge vertices are shown in Fig.~\ref{fig:zeroAndDoubleChargeModes} a) and b), respectively. These mode profiles are complementary in the sense that only double-charge vertices are active at \SI{7.9}{\giga\hertz}, and only the zero-charge vertices participate in the mode at \SI{14.2}{\giga\hertz}. In spite of this clear separation in terms of frequency and micromagnetic configuration, the oscillation amplitude of the different vertices is not identical within each of these modes. Some vertices display strong oscillations, whereas a few vertices show only little activity. We attribute this inhomogeneity in the amplitude to local magnetic fields within the array arising from the charged vertices. The frequency of the zero-charge vertices and the local mode profile of the active vertices remain mostly the same as in the ordered AMC, but the phase correlation between the vertices is lost in the disordered state (see Movie 6 of the supplemental material).
Compared with the ordered case, the intensity of the zero-charge vertex mode \fbox{2} is reduced in the disordered state, in accordance with the lower density of this type of vertices in the disordered AMC.

A more significant difference in the spectra concerns the peaks \fbox{3} and \fbox{4} describing the first and the second standing-spin-wave modes in the nanowires, which significantly broaden and become almost indiscernible in the disordered AMC. Instead of the two well-defined peaks, a weakly defined quasi-continuum of overlapping modes appears in the frequency range between 15 and \SI{19}{\giga\hertz}. A detailed Fourier analysis of the individual nanowires shows that these modes are still present in the disordered crystal. Their frequency, however, differs from one nanowire to another, resulting in a broadened and blurred spectrum in that frequency range. This effect, too, can be explained by spatial variations of local magnetic fields in the disordered state.

Charged vertices generate local magnetic fields within the disordered crystal that are absent in the ordered state. The magnetic divergences developing, for example, in a vertex with negative double charge (three-in/one-out configuration) and those in a positive double charge vertex (one-in/three-out) are monopole-type sources and sinks of the magnetic field $\vec{H}$, respectively. Correspondingly, a longitudinal magnetic field develops in a nanowire forming a Dirac string that connects two such vertices of opposite charges \citep{mengotti_real-space_2011, gliga_spectral_2013}. Such longitudinal magnetic fields result in a shift of the standing-wave modes to different frequencies \cite{jorzick_brillouin_1999}.
There are numerous possible combinations of vertices that can be pairwise connected by a nanowire in the disordered state since vertices can differ in both magnitude and sign of the magnetic charge they carry. Accordingly, although the homogeneous and axial magnetization structure is the same in all nanowires, their high-frequency magnetic dynamic properties differ as a result of a broad distribution of local fields throughout the crystal. In addition to differences in magnitude, the magnetic field developing in wires connecting oppositely charged vertices can also differ in sign. The dipolar magnetic field generated by the vertex charges can be parallel or antiparallel to the wire magnetization. This leads to two qualitatively different cases, resulting in an additional lift of the degeneracy of nanowire modes.  
Compared to the modes \fbox{4} and \fbox{5}, the higher-frequency excitation \fbox{6} at \SI{28}{\giga\hertz} remains largely unaffected by the magnetic disorder of the crystal. In general, high-frequency modes are dominated by the exchange interaction, which leads to local effective fields that are much larger than those due to dipolar fields. Correspondingly, changes in the dipolar field distribution have a weaker impact on exchange-dominated high-frequency modes.

\section{Impact of structural defects}

\begin{figure}[ht]
\includegraphics[width=\linewidth]{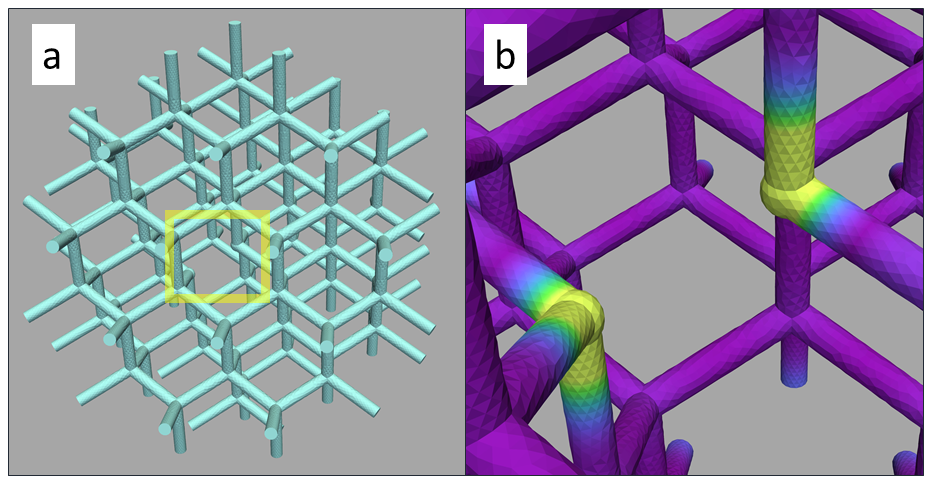}
\caption{Structural defect within the AMC giving rise to an additional mode localized at the defect site. Frame (a) shows the modified AMC in which a single nanowire has been removed. The removal of a nanowire introduces two vertices with reduced coordination number which oscillate at \SI{11.7}{\giga\hertz}. The spatial distribution of this mode is localized at the defect site (b).  
This defect mode appears in the spectrum as a small peak labeled \fbox{7} in the inset of Fig.~\ref{fig:defectFreeSpectrum} 
\label{fig:defectMode}.}
\end{figure}

Having established that charged vertices, which represent {\em magnetic} defects, strongly affect the high-frequency properties of the AMC, we now analyze how a {\em structural} defect changes the magnetic spectrum. We introduce a structural defect by removing a single nanowire inside the AMC, as shown in Fig.~\ref{fig:defectMode}a). This removal results in two defect vertices with reduced coordination number as they represent the junction of three instead of four nanowires. To isolate the effect of this structural defect on the high-frequency properties of the AMC, it is introduced as a modification of the otherwise defect-free magnetic state, i.e., the ordered reference state. Despite obeying the spin-ice rule, both defect vertices have a charged magnetic structure. With their two-in/out-out and one-in/two-out type structures, the defects display a magnetic charge of type $\pm 1q$ while preserving the overall charge neutrality of the AMC. Although they have a magnetostatic charge equal to that of surface vertices, the defect vertices are different in terms of their magnetic structure and charge type. The defect vertices carry volume-type magnetostatic charges, while the free ends have surface-type charges \cite{brown_magnetostatic_1962}.
The magnonic spectrum of the AMC with a structural defect [represented as line (c) in Fig.~\ref{fig:defectFreeSpectrum}] demonstrates the appearance of an additional, weakly pronounced mode (label \fbox{7} in the inset). As shown in Fig.~\ref{fig:defectMode}b), this mode is localized at the defect vertices. It has a low intensity in the spectrum because the signal originates from only two defect vertices out of a total of more than 100 surface and volume vertices in the AMC. Besides this additional low-amplitude peak, the introduction of a structural defect also leads to a clear shift of the volume modes \fbox{3}, \fbox{4}, and \fbox{5} toward lower frequencies. Adding more structural defects exacerbates this effect and leads to further changes in the spectrum. Upon increasing the density of structural defects, the extent to which the spectrum changes quickly reaches a degree at which it becomes impossible to consider the defects as mere perturbations or modifications of the defect-free state.

\section{Field dependence of magnonic spectra}
To characterize the magnonic properties of the AMC more broadly, we investigate how the absorption spectra evolve when an external magnetic field is applied. 
\begin{figure}[ht]
\includegraphics[width=\linewidth]{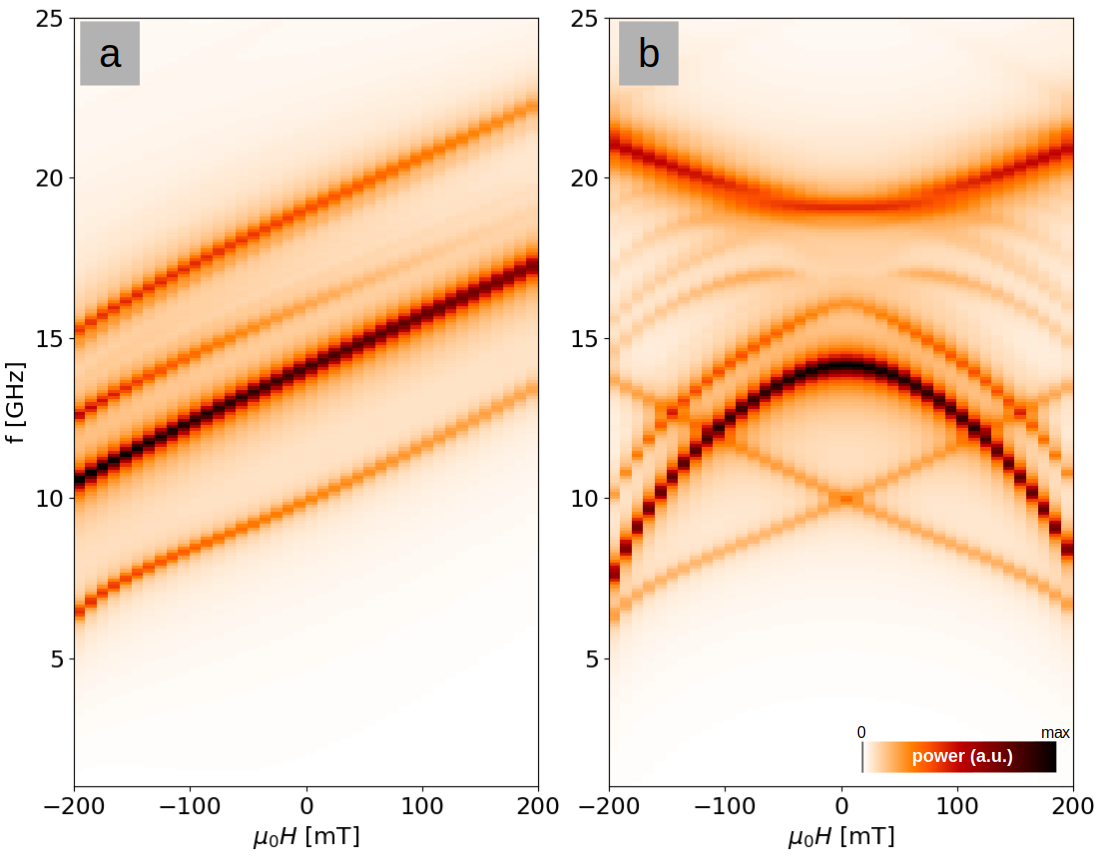}
\caption{Field-dependence of the magnonic absorption spectra of the defect-free, high-remanence configuration when a magnetic field is applied parallel (a) and perpendicular (b) to the remanent magnetization direction. 
\label{fig:fieldspectra} }
\end{figure}
For simplicity, we consider the field-dependence of the spectrum of the ordered configuration (a) in Fig.~\ref{fig:defectFreeSpectrum}, which is a defect-free state in terms of geometry and magnetic structure. This high-remanence state is obtained through quasi-static reduction of an external field from saturation to zero. The field is applied along the $[001]$ crystallographic direction of the AMC's unit cell, which corresponds to the $z$ axis in Fig.~\ref{fig:projections}. In this case, all four sets of nanowires enclose the same angle with the external field. 
We denote this field direction as parallel (to the crystal's remanent magnetization), to contrast it with the other situation we study (perpendicular), where we apply the field in the [100] direction, i.e., at an angle orthogonal to the average magnetization at remanence.

In both cases, we vary the field strength in a range of $\pm$\SI{200}{\milli\tesla} in steps of \SI{10}{\milli\tesla}. At each field value, we calculate the static magnetization structure and its oscillatory small-amplitude dynamics.
Figure ~\ref{fig:fieldspectra} displays the spectra of the parallel (a) and the perpendicular case (b). The data is obtained using a newly developed numerical method \cite{daquino_micromagnetic_2023} specifically tailored to simulate such situations. The intensity of the color code displays the absorbed power at different field strengths and frequencies, with the zero-field line reproducing the first four peaks shown in the spectrum (a) of Fig.~\ref{fig:defectFreeSpectrum}.

In the parallel case, shown in Fig.~\ref{fig:fieldspectra}a, the resonance frequencies vary essentially linearly with the applied field strength, with a slope of about \SI{18}{\giga\hertz\per\tesla}. 
In the considered field strength, the intensity of the individual resonances is mainly independent of the applied field, except for the third peak (denoted as \fbox{3} in Fig.~\ref{fig:defectFreeSpectrum} and identified as the first standing-wave mode within the nanowires), whose intensity decreases with positive and increases with negative field strength. The field-dependence of the absorption spectrum changes significantly if the field is applied perpendicular to the direction of remanence, as shown in Fig.~\ref{fig:fieldspectra}b. In this case, the resulting magnonic band structure is symmetric about the $H=0$ axis and displays various features. The lowest-frequency mode \fbox{1}, associated with the oscillations of the free ends at the surfaces, experiences a Zeeman-type splitting. In contrast, the other modes show a more complex field dependence, with some displaying a decrease and others increasing their frequency with increasing (absolute) field strength. Moreover, additional absorption lines appear at values $|\mu_0H_\text{ext}|>\SI{100}{\milli\tesla}$, some of which appear to branch out from the second-order spin wave mode \fbox{4} as the field strength increases. The field-dependent absorption spectra are thus strongly anisotropic: an external field can modify the spectrum very differently depending on whether it is applied perpendicular or parallel to the remanence direction.

\section{Conclusion}
We have identified a wide variety of high-frequency magnetic modes in an AMC consisting of a three-dimensional network of interconnected nanowires arranged in the form of a diamond-type artificial crystal structure. The magnetic modes of the AMC can be classified into two categories: the fundamental excitations, which are related to the geometrical constituents of the network (vertices, nanowires, surfaces), and the defect modes, which are linked to the magnetic structure of the AMC. Magnetically charged vertices within the crystal give rise to distinct additional modes and yield unmistakable fingerprints in the magnonic spectrum. These charged vertices, which represent magnetic defects in the terminology of artificial spin ice, thus lead to effects analogous to the changes that atomic defect sites induce in the optical spectrum of natural diamonds \cite{maki_properties_2009,aharonovich_diamond_2014}. We anticipate that our findings, which demonstrate that high-frequency modes can be controlled through both the 3D geometry and the magnetic structure of an AMC, will further promote three-dimensional interconnected nanowire arrays as a new category of magnonic metamaterials \cite{kadic_3d_2019} with promising potential for applications in devices for storage, magnonic operations, and neuromorphic computations \cite{grollier_neuromorphic_2020, skjaervo_advances_2020}.

\section{Methods}

\subsection{Micromagnetic model}
The magnetic structure and the dynamic modes are calculated using our custom-developed micromagnetic finite-element software {\tt tetmag}, \cite{hertel_tetmag_2023, hertel_large-scale_2019}. Given an initial magnetic configuration $\vec{M}(\vec{x})$, the algorithm integrates the Landau-Lifshitz-Gilbert equation \cite{landau_theory_1935,gilbert_phenomenological_2004}
\be\label{llg}
\frac{d\vec{M}}{dt}= -\gamma_0\left(\vec{M}\times\vec{H}_\text{eff}\right)
+\frac{\alpha }{M_s} \left(\vec{M}\times\frac{d\vec{M}}{dt}\right)
\ee
in time and thereby yields information on the structure of the magnetization and its dynamics. The effective field $\vec{H}_\text{eff}$ is defined by the variational derivative of the micromagnetic energy $E$ with respect to the magnetization, $\mu_0\vec{H}_\text{eff} = -\delta E(\vec{M})/\delta\vec{M}$, where $\mu_0=4\pi\times 10^{-7}\si{\volt\second\per\ampere\per\meter}$. The total energy $E=E_\text{exc}+E_\text{zee}+E_\text{mag}$ is the sum of the exchange energy $E_\text{exc}=-M_s^{-2}\int A\cdot\vec{M}\Delta\vec{M}\dv$, the Zeeman energy $E_\text{zee}=-\mu_0\int\vec{M}\cdot\vec{H}_\text{ext}\dv$, and the magnetostatic energy $E_\text{mag}=(\mu_0/2) \int \vec{M}\cdot\nabla u\dv$, where $A$ is the ferromagnetic exchange constant, $\vec{H}_\text{ext}$ is an externally applied field, and $u$ is the magnetostatic potential \cite{brown_magnetostatic_1962}. 
In equation (\ref{llg}), $\alpha$ is a dimensionless damping constant and $\gamma_0/2\pi = \SI{28.02}{\giga\hertz\per\tesla}$ is the gyromagnetic ratio, and $M_s = \left|\vec{M}\right|$ is a material-specific constant; the saturation magnetization. The magnetic material of this study is described by the parameters of FEBID-deposited cobalt: $\mu_0M_s=\SI{1.2}{\tesla}$, zero magnetocrystalline anisotropy, and $A=1.5 \times 10^{11}\si{\joule\per\meter}$ \cite{AmalioCommunication}.

The numerical calculation uses a discretized representation of the magnetization in which the vector field $\vec{M}$ is defined at the nodes of a tetrahedral finite element mesh. We use netgen \cite{schoberl_netgen_1997} to generate the finite-element meshes and made sure that the element size remained everywhere below the magnetostatic exchange length $l_s =\sqrt{2\mu_0A/M_s^2}=\SI{5.1}{\nano\meter}$. In this study, the choice of the maximum cell size was primarily imposed by geometrical aspects rather than by the intrinsic material parameters. To accurately reproduce the nanometric rounded features of the modeled shape, the discretization size was set to $\SI{2.5}{\nano\meter}$ -- significantly smaller than $l_s$.

\subsection{Preparation of zero-field magnetic states}
Stable minimum-energy configurations of the magnetization are calculated by numerically integrating the Landau-Lifshitz-Gilbert (LLG) equation in time. Convergence is reached when the magnitude of the local torque $\left|\vec{M}\times\vec{H}_\text{eff}\right|$ drops below a threshold value at every discretization point. In this type of LLG-based energy-minimization, the magnetization dynamics towards the relaxed state is unimportant and therefore, to accelerate the calculation, the damping parameter $\alpha$ can be set to an unrealistically large value. We typically use $\alpha=0.5$ in these cases.

The defect-free reference state of the AMC is obtained by starting from a homogeneous magnetization state and relaxing the system at zero applied field. The disordered state, on the other hand, is prepared by minimizing the system's total energy at zero field after starting from a random initial configuration. 

\subsection{Mode calculation}
The high-frequency modes are generated by applying a small perturbation to the relaxed state---typically a low-amplitude picosecond magnetic field pulse---and analyzing the magnetization dynamics that it induces. It is important to choose a perturbation that is, on one hand, sufficient to generate a dynamic response while, on the other hand, avoiding magnetic switching processes or other persistent modifications of the zero-field configuration. When a suitable perturbation is applied, it triggers a characteristic ``ringing'' of the magnetization arising from the small-angle precession dynamics of the local magnetization around the equilibrium state. 
This ringing fades out after a few nanoseconds as the system asymptotically returns to the converged state. To extract the magnetic modes, the oscillations during this ringdown process are recorded and Fourier-analyzed~\cite{cheenikundil_switchable_2021,gliga_spectral_2013,yan_calculations_2007}. The overall spectrum, as shown in Fig.~\ref{fig:defectFreeSpectrum}, is the sum over all discretization points of the Fourier-transformed magnetic oscillation. A low value of the damping constant $\alpha$ must be chosen to ensure a sufficiently large number of oscillations and narrow line width in the Fourier spectrum; we have set $\alpha = 0.01$ in these dynamic calculations. 

The profiles of the individual modes are filtered from the Fourier-transformed magnetic oscillations by means of a windowed Fourier back-transform from the frequency to the time domain. This Fourier back-transformation is performed such as to only consider the oscillations occurring within a narrow frequency range near the frequency of interest \cite{yan_calculations_2007, baker_proposal_2017}. In addition to this windowing in the frequency domain, we also sometimes restrict the analysis to certain spatial regions, for instance when we analyze the modes unfolding within individual nanowires.
\section*{Acknowledgments}
This work was funded by the LabEx NIE (ANR-11-LABX-0058\_NIE) in the framework of the Interdisciplinary Thematic Institute QMat (ANR-17-EURE-0024), as part of the ITI 2021-2028 program supported by the IdEx Unistra (ANR-10-IDEX-0002-002) and SFRI STRATUS (ANR-20-SFRI-0012) through the French Programme d'Investissement d’Avenir.
RC and RH acknowledge the High Performance Computing center of the University of Strasbourg for supporting this work by providing access to computing resources. 
\bibliography{diamond.bib}      
\end{document}